\documentclass[
reprint,
amsmath,amssymb,
aps,
footinbib
]{revtex4-1}

\usepackage{graphicx}% Include figure files
\usepackage{dcolumn}% Align table columns on decimal point
\usepackage{bm}% bold math
\usepackage{hyperref}% add hypertext capabilities
\usepackage{csquotes}
\usepackage{gensymb}
\hypersetup{colorlinks=true,linkcolor=red,citecolor=red}
\usepackage{float}
\usepackage[dvipsnames]{xcolor}

\begin{document}

\title{Impact of chemical disorder on magnetic exchange interactions in L1$_0$-FeNi (tetrataenite)}

\author{Ankit Izardar}
\author{Claude Ederer}
\affiliation{Materials Theory, ETH Z\"urich, Wolfgang-Pauli-Strasse 27, 8093 Z\"urich, Switzerland}

\date{\today}

\begin{abstract}
We investigate the effect of chemical disorder on the magnetic exchange couplings and the Curie temperature ($T_{\text{c}}$) in L1$_0$-ordered FeNi using first-principles-based calculations. We use supercells to model chemical disorder, to account for the specific symmetry-broken local chemical environments around the individual atoms. We find a very strong variation of the most dominant first-nearest neighbor Fe-Fe interaction for different inequivalent Fe-Fe pairs, ranging from around 5\,meV to 37\,meV, compared to a coupling strength of 27\,meV in the ordered state. To estimate the influence of such strong variations of the magnetic coupling constants on the Curie temperature of the disordered or partially ordered state, we study a simple Heisenberg model with random Gaussian-distributed nearest neighbor couplings on an fcc lattice. Our Monte Carlo simulations for this model indicate that strongly varying exchange couplings, such as those obtained for FeNi, can lead to a reduction of $T_{\text{c}}$ of around 10\,\% relative to the one obtained using only the average coupling.

\end{abstract}

\maketitle

\section{\label{sec:Intro}Introduction}

The chemically ordered ferromagnet L1$_0$-FeNi (tetrataenite) has recently generated considerable interest as a rare-earth-free, low-cost permanent magnet, due to its high magneto-crystalline anisotropy energy and large saturation magnetization~\cite{Paulev_1968, kojima_2014, Lewis2014, Lewis2014InspiredBN, Alex, Werwi_ski_2017, CUI2018118, Tian2019, Tian2020, Izardar_2020, Tian2021}. Since its discovery by N{\'e}el and coworkers in the early 1960s~\cite{Pauleve1962,Neel1964}, several attempts have been made to synthesize L1$_0$-FeNi with a high degree of chemical order~\cite{Nitrogen-insertion, Amorphous-to-crystalline, PLD, Fe-Ni-composition}. However, the synthesis of a fully ordered structure remains challenging, due to the rather low order-disorder transition temperature, which is around 590\,K~\cite{Pauleve1962, Neel1964, Reuter1989}. At this temperature, the diffusivity of atoms is too low for the ordered structure to form on reasonable timescales. Therefore, ``naturally occurring'' tetrataenite has only been found in iron meteorites~\cite{ALBERTSEN1978, PETERSEN1977192, Danon1979, Danon1980, clarke}. 
Due to the difficulties in obtaining fully ordered samples, it becomes essential to investigate and understand how deviations from the perfect order affect the magnetic properties, in particular the Curie temperature, $T_{\text{c}}$, and magnetic anisotropy, of L1$_0$-FeNi. 

Experimentally, only the Curie temperature of the disordered system ($T_c \approx 785$\,K~\cite{Onodera1981,Wei2014}) is accessible, since the ordered system disorders on heating at temperatures above $\sim$ 700\,K. Thereby, the effective ``disordering temperature'' depends strongly on the heating rate~\cite{DOSSANTOS2015234}. On time-scales typical for magnetization measurements, disordering occurs around 820\,K, i.e., above the Curie temperature of the disordered system, and results in an abrupt vanishing of the magnetization~\cite{Lewis2014}. This indicates that the nominal Curie temperature of the ordered system would be noticeably higher than that of the disordered system. 

Several previous studies have used first-principles calculations to obtain the Curie temperature in L1$_0$-FeNi. For example, Edstr\"om {\it et al.}~\cite{Alex} and Tian {\it et al.}~\cite{Tian2020} obtained values for the Curie temperature of the ordered phase of 916\,K and 780\,K, respectively, using slightly different electronic structure methods. They also found that chemical disorder leads to a reduction of the Curie temperature. 

Both of these studies have used the coherent-potential approximation (CPA) \cite{CPA1, CPA2} to incorporate chemical disorder in the material. The CPA is based on an effective medium description of the atomic environments, and thus provides a very efficient method for the treatment of disorder effects in random alloys using only a single unit cell. However, CPA does not include effects related, e.g,. to the local symmetry-breaking of a specific chemical environment around an individual atom.
Such effects beyond CPA can be particularly relevant, e.g., for the magnetic anisotropy, as we showed in our previous work, where we have used supercells with different distributions of Fe and Ni atoms to investigate the effect of chemical disorder~\cite{Izardar_2020, Si_2021}. These calculations indicate that, for example, a moderate increase in Fe content, while reducing the degree of chemical order in the system, leads to an increase of the magneto-crystalline anisotropy energy, an effect not captured within the CPA. 

In the present work, we use first principles calculations based on density functional theory (DFT) to provide further insights into the effect of variations in the local chemical environment on the magnetic exchange interactions, and consequently the Curie temperature, in partially ordered FeNi. To model the chemical disorder, we follow a similar approach as in our previous work~\cite{Izardar_2020, Si_2021}, i.e., we incorporate the effect of a disordered local atomic environment on the magnetic coupling by employing supercells. 
We find that treating the disorder on a local level gives rise to remarkably strong variations of the first nearest neighbour Fe-Fe coupling. 
Our subsequent analysis of the correlation between the specific local chemical environment and the corresponding magnetic exchange couplings indicates that the magnetic coupling is governed by long range effects that clearly go beyond the closest neighbor environment. This makes it extremely challenging to consider such configuration-dependent couplings for the calculation of $T_{\text{c}}$ and other thermodynamic properties.
In order to obtain a rough estimate of how such strong variations of the magnetic coupling constants will affect the Curie temperature, compared to using only an average coupling, obtained, e.g. from an effective medium treatment of chemical disorder,  we perform Monte Carlo simulations for a simple Heisenberg model with random Gaussian-distributed coupling constants. We find that variations of the same order as obtained in our DFT calculations for FeNi can lead to a reduction of $T_c$ of around 10\,\%.

In the following, we first describe the computational method we use to obtain magnetic exchange couplings, and then present our results for both ordered and partially disordered FeNi.

\section{\label{sec:computational_method}Computational method}

\subsection{\label{subsec:MFT}Magnetic exchange interactions}

Within the (classical) Heisenberg model, the energy of a magnetic system is expressed as a sum over pairwise (bilinear) interactions between localized magnetic moments:
\begin{equation}\label{eq:heisenberg_hamilt}
E = -\frac{1}{2}\sum_{i\neq j}J_{ij}\textbf{S}_i \cdot \textbf{S}_j \quad .
\end{equation}
We use the convention that $\textbf{S}_i$ is a normalized vector describing only the direction of the magnetic moment at site $i$. 

It is well known that, for an itinerant magnetic material such as FeNi, the Heisenberg model is not necessarily a good approximation \cite{book_spin_Moriya}, and Eq.~\eqref{eq:heisenberg_hamilt} is typically only valid for not too large fluctuations around the ferromagnetic ground state.
Thus, to calculate magnetic exchange couplings, $J_{ij}$, we use the following well-known equation based on the magnetic force theorem, which is obtained by considering the energy variation with respect to infinitesimal rotations of the magnetic moments~\cite{LIECHTENSTEIN198765}:
\begin{widetext}
\begin{equation}
\label{eq:Jij_MFT}
    J_{ij} = \frac{1}{2\pi}\mathrm{Im}\int_{-\infty}^{\varepsilon_\mathrm{F}}\mathrm{d}\varepsilon\sum_{mm'm''m'''}\Delta_{i}^{mm'}G_{ij,\downarrow}^{m'm''}(\varepsilon)\Delta_{j}^{m''m'''}G_{ji,\uparrow}^{m'''m}(\varepsilon) \quad .
\end{equation}
\end{widetext}
Here, $\Delta_{i}^{mm'}$ ($\Delta_{j}^{m''m'''}$) is the local exchange splitting on site $i$ ($j$), and $G_{ji,\uparrow}^{m'''m}$ ($G_{ij,\downarrow}^{m'm''}$) is the spin-up (spin-down) intersite Green's function. Both quantities are expressed within a tight-binding-like basis, where each basis-orbital (with index $m$) is localized on a specific site.

To evaluate the quantities in Eq.~\eqref{eq:Jij_MFT}, we first obtain the electronic structure from plane-wave-based density-functional theory (DFT) calculations, and then transform the corresponding Kohn-Sham Hamiltonian into a basis of localized Wannier functions~\cite{Rudenko_et_al:2013,Korotin_et_al:2015,Logemann_et_al:2017,SrMnO3-Xiangzhou}.
As described in more detail in Sec.~\ref{subsec:WF}, we use Wannier functions defined by orbital projection and subsequent orthonormalization (corresponding to the ``initial projections''  in the {\tt wannier90} code~\cite{MOSTOFI20142309}).
This leads to a set of atom-centered basis orbitals. As shown in Sec.~\ref{subsec:WF}, an excellent representation of all occupied bands in FeNi can be achieved by using a full set of $s$, $p$, and $d$ projections for each atom. 

In contrast, constructing maximally localized Wannier functions (MLWFs)~\cite{MLWFs_1997} for FeNi, results in a set of Wannier functions where the Wannier orbitals corresponding to the $s$ and $p$ projections become localized in between the atoms, and thus cannot be used to evaluate Eq.~\eqref{eq:Jij_MFT}. This is similar to what has been described for the nearly free-electron-like bands in fcc Cu (and other $3d$ transition metals), see e.g., Ref.~\onlinecite{Souza/Marzari/Vanderbilt:2001}. Further details are presented in Sec.~\ref{subsec:WF}.

After a suitable set of Wannier functions has been constructed, we follow the approach outlined in Ref.~\cite{SrMnO3-Xiangzhou} to obtain exchange couplings for different pairs of atoms. 

\subsection{\label{sec:MFT}Computational details}

In order to accommodate both the fully ordered L1$_0$ structure of FeNi as well as some configurations with (partial) chemical disorder, we use an 8-atom cell, corresponding to a $\sqrt{2} \times \sqrt{2} \times 1$ supercell of the conventional cubic 4-atom fcc unit cell, or, equivalently,  to a $2 \times 2 \times 1$ supercell of the 2-atom tetragonal primitive unit cell of the L1$_0$ structure. 
For test purposes, we also perform some calculations for the perfectly ordered L1$_0$ structure using the 2-atom primitive unit cell.
In all cases, except where otherwise noted, we fix the lattice parameters and atomic positions to that of a perfectly cubic fcc lattice with lattice constant $a=3.56$\,\AA, and then distribute Fe and Ni atoms over the available sites within the cell in different ways.

We perform DFT calculations using the Vienna \textit{ab} \textit{initio} Simulation package (VASP) \cite{Kresse1996}, the projector-augmented wave method (PAW) \cite{PAW1994,Kresse1999}, and the generalized gradient approximation according to Perdew, Burke, and Ernzerhof (PBE)~\cite{PBE}. Brillouin zone integrations are performed using the tetrahedron method with Bl\"{o}chl corrections and a $\Gamma$-centered $12\times12\times16$ \textbf{k}-point mesh for the 8-atom cell. The plane wave energy cut-off is set to 550\,eV, and the total energy is converged to an accuracy of $10^{-8}$\,eV. Our PAW potentials include $3p$, $4s$, and $3d$ states in the valence for both Fe and Ni. All calculations are performed for the ferromagnetically ordered state. 

A Wannier representation of the Kohn-Sham Hamiltonian is then obtained using the {\tt wannier90} code \cite{MOSTOFI20142309}, using the same $k$-point mesh as for the DFT calculations.
To check the convergence of the calculated magnetic exchange couplings with respect to the \textbf{k}-point sampling, we perform calculations using up to $14\times14\times18$ \textbf{k}-points and find our results to be sufficiently converged using a $12\times12\times16$ \textbf{k}-point mesh.

To obtain the Curie temperature for the ordered case, using the Heisenberg model, Eq.~\eqref{eq:heisenberg_hamilt}, with the coupling constants obtained from our DFT calculations, we perform Metropolis Monte Carlo simulations, as implemented in the UppASD package~\cite{Skubic_2008}. 
We consider magnetic exchange couplings for all pairs of atoms up to a distance of 10$a$ ($\sim$ 25.17 \AA{}). To accurately determine $T_{\text{c}}$, accounting for potential finite size effects due to the limited size of our simulation cells, we use the Binder cumulant method~\cite{landau_binder_2014}. Thus, $T_{\text{c}}$ is obtained as the temperature where the fourth-order Binder cumulants, obtained for three different cell sizes, cross. We consider cell-sizes of $20 \times 20 \times 20$, $30 \times 30 \times 30$, and $36 \times 36 \times 36$, relative to the primitive tetragonal 2-atom cell of the L1$_0$ structure.
For the test calculations presented in Fig.~\ref{img:Tc_vs_distance}(b), we use a cell size of $20 \times 20 \times 20$. 

For the model study presented in Sec.~\ref{subsec:model_study}, we use a $15 \times 15 \times 15$ supercell of the conventional fcc cubic cell. We then initialize the magnetic couplings for all nearest neighbor pairs within this cell individually by drawing random numbers from a Gaussian distribution with varying standard deviation, $\sigma>0$, and a mean value of $\mu=1$. 
For $\sigma = 0.0$, the coupling constants of all nearest neighbor pairs are identical and equal to the mean of the Gaussian distribution ($\mu = 1.0$).
The Curie temperature for each $\sigma > 0.0$ is obtained by taking an average over 100 instances of the Gaussian-distributed magnetic exchange couplings with different random seeds.  
Convergence with respect to the number of instances of the randomized system has been verified 
by monitoring the cumulative average of $T_{\text{c}}$ with increasing number of instances over all 100 samples.

\section{\label{sec:results}Results and Discussion}

\subsection{\label{subsec:WF} Construction of Wannier functions}

\begin{figure}
   \centering
   \includegraphics[width=\columnwidth]{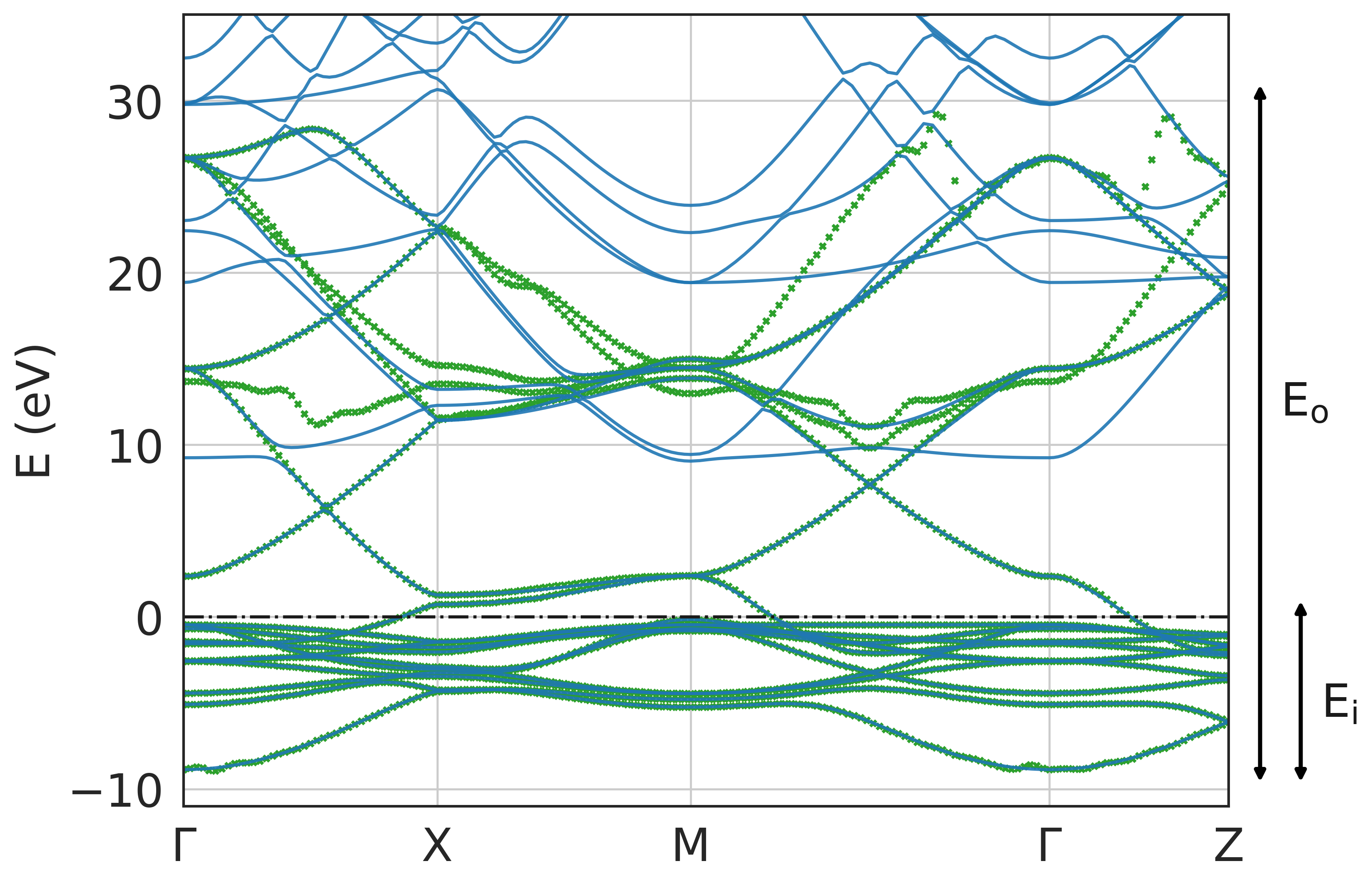}
   \caption{Majority spin bandstructure of chemically ordered L1$_0$-FeNi. Blue and green dots represent the Kohn-Sham and Wannier-interpolated bands, respectively. Outer and inner energy windows, $E_\text{o}$ and $E_\text{i}$, are indicated on the right side of the plot. The Fermi level defines zero energy.}
   \label{fig:bandstructure_2atom_cell}
\end{figure}

As outlined in Sec.~\ref{subsec:MFT}, we start by constructing a set of  Wannier functions from orthonormalized projections on $s$, $p$, and $d$ orbitals for each Fe and Ni atom in the unit cell. We use an outer energy window ranging from $-10$\,eV up to about 31\,eV, which contains all occupied valence bands plus a certain number of empty bands. Furthermore, in order to accurately reproduce all occupied bands, we employ a frozen (inner) energy window from $-10$\,eV up to about 1\,eV above the Fermi energy. The band dispersion (only for the majority spin component) obtained from the resulting Wannier functions for the minimal 2-atom cell of the fully ordered L1$_0$ structure is shown in Fig.~\ref{fig:bandstructure_2atom_cell}, together with the underlying Kohn-Sham bandstructure. 

One can see that all bands below $\sim$ 10\,eV are well described by the Wannier-interpolated bands. Some weak oscillations can be seen in the lowest lying, free-electron-like band around the $\Gamma$-point. These are due to the fact that the $k$-mesh along the high-symmetry lines used to obtain the interpolated bandstructure is much finer than the homogeneous $k$-mesh used to construct the Wannier functions, and that, in order to obtain atom-centered Wannier functions, we do not apply the usual ``disentanglement procedure'' to obtain an optimally $k$-connected subspace. 
% \CE{Do these oscillations vanish if we use disentanglement/maximum localization?} \AI{Yes.}
We note that the calculation of the magnetic exchange couplings is based on the original homogeneous $k$-point mesh, where the Wannier bands are identical to the DFT Kohn-Sham bands by construction.
For the minority-spin bands (not shown), we obtain a similar good agreement between the Wannier-interpolated and the occupied Kohn-Sham bands.

\subsection{\label{subsec:ferro} Magnetic interactions in L1$_0$-ordered FeNi}

\begin{figure}
   \centering
   \includegraphics[width=\columnwidth]{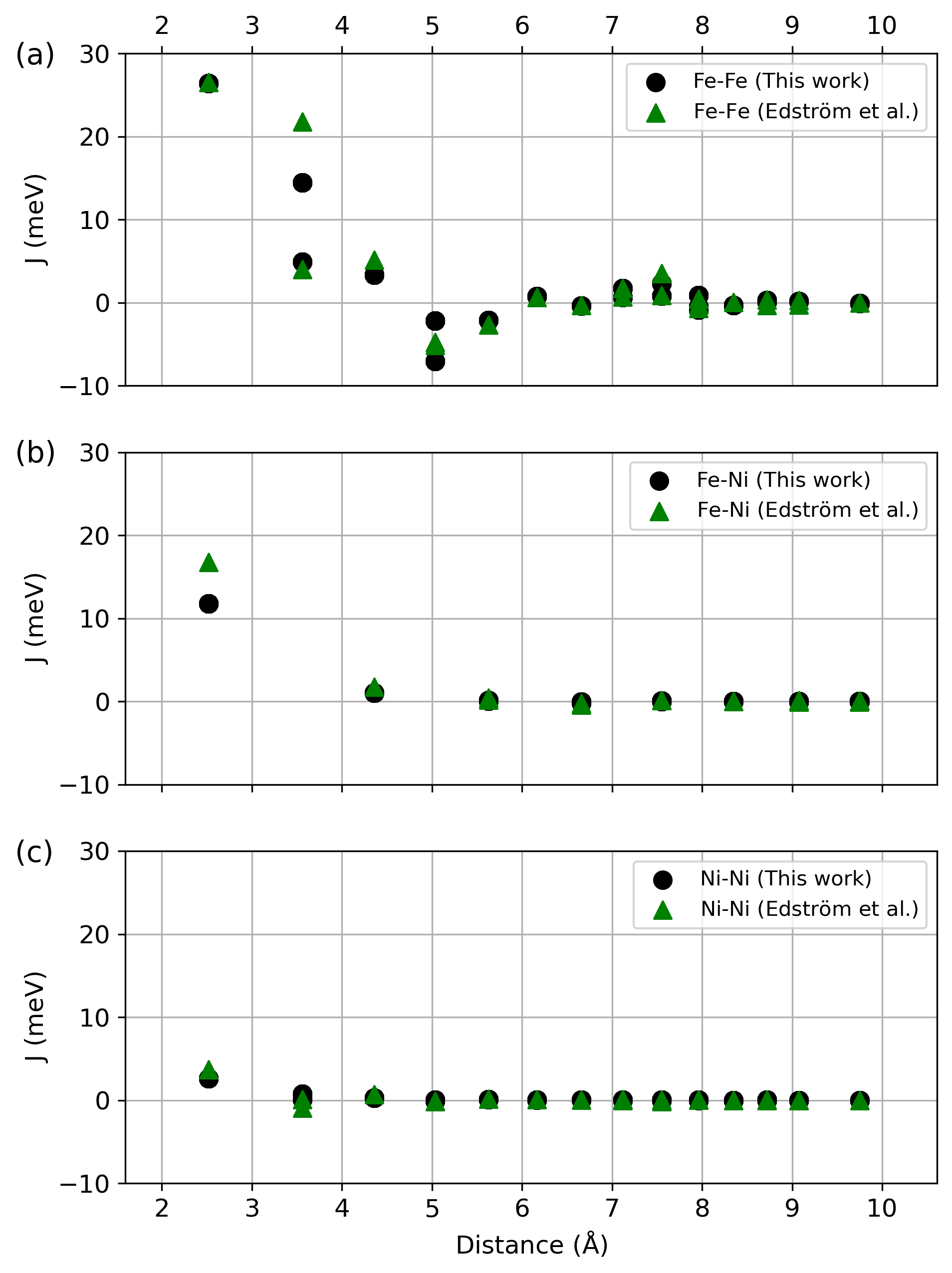}
   \caption{Magnetic exchange couplings for (a) Fe-Fe, (b) Fe-Ni, and (c) Ni-Ni pairs as a function of atomic distance in the ordered L$1_0$ phase of FeNi as calculated in this work (black dots) and by Edstr\"om et al.~\cite{Alex} (green triangles).}
   \label{img:J_vs_d_comparison_Alex}
\end{figure}

Next, we calculate the magnetic exchange couplings for the ordered L$1_0$-phase of FeNi, using the method described in Sec.~\ref{subsec:MFT}. In Fig.~\ref{img:J_vs_d_comparison_Alex}, we compare the magnetic exchange couplings obtained in the present study (shown by the black dots) with those calculated by Edstr\"om \textit{et al.}~\cite{Alex} 
% using the spin-polarized relativistic Korringa-Kohn-Rostocker method 
(shown by the green triangles). The different sub-panels show the couplings corresponding to Fe-Fe, Fe-Ni, and Ni-Ni pairs as a function of distance. 
Overall, there is very good agreement between the two data-sets, except for a few cases discussed further below. One can see that the Fe-Fe couplings are strongest and rather long-ranged while the Fe-Ni and Ni-Ni interaction is weaker and decays rather quickly with distance. 

Note that for certain distances (e.g., for the one corresponding to second nearest-neighbour distance, $d_{ij}=a_0=3.56$\,\AA), two distinct values for $J_{ij}$ are obtained, depending on the orientation (in-plane versus out-of-plane) of the corresponding pair relative to the tetragonal axis, i.e., the axis defined by the long range order. In particular for the second nearest neighbor Fe-Fe interaction, this difference is rather large. This already indicates a strong configuration dependence of the magnetic coupling, which will be further analyzed in Sec.~\ref{subsec:couplings_disordered} using supercells with partial chemical disorder. 

For the second nearest neighbor Ni-Ni pairs, the coupling is rather weak and thus the (absolute) difference between in-plane and out-of-plane coupling is small.  
Furthermore, one can see that Fe-Ni couplings for certain distances are ``missing'' (e.g., corresponding to second nearest neighbors on the fcc lattice, $d_{ij}=a_0=3.56$\,\AA). This is due to the arrangement of Fe and Ni atoms in the underlying L$1_{0}$ structure. 

The good agreement between our results and the calculations of Edstr\"om \textit{et al.} (Ref.~\onlinecite{Alex}) is remarkable, since rather different electronic structure methods, involving different approximations and basis sets, have been employed.
Furthermore, our calculations are based on a metrically cubic fcc lattice, whereas Edstr\"om \textit{et al.} have used tetragonal lattice vectors with a slightly different $c/a$ ratio. 
The biggest difference is observed for the second nearest neighbor (in-plane) Fe-Fe coupling and the first nearest neighbor Fe-Ni coupling, for which Edstr\"om \textit{et al.} obtain a noticeably stronger coupling (by around 7 meV and 5 meV, respectively). Our calculated coupling constants also appear to be in good agreement with more recent calculations by Tian \textit{et al.}~\cite{Tian2020}~\footnote{Note that Tian \textit{et al.} use a different normalization in the Heisenberg model and therefore the absolute values in Ref.~\onlinecite{Tian2020} are different from ours.}  

\begin{figure}
   \centering
   \includegraphics[width=\columnwidth]{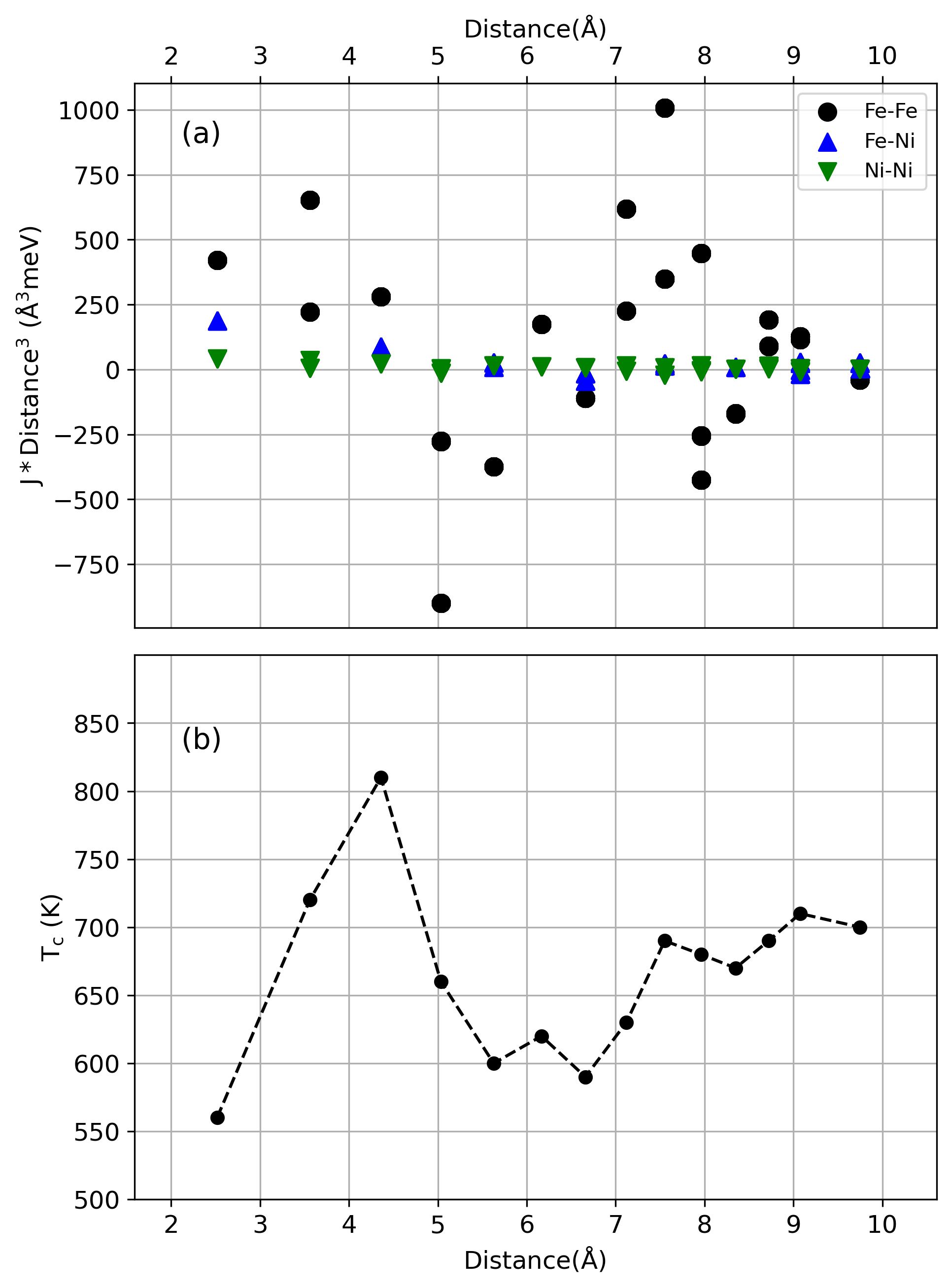}
   \caption{(a) Magnetic coupling constants as a function of atomic distance multiplied with the cube of the corresponding atomic distance, $J_{ij}\cdot d_{ij}^3$, calculated for the ordered phase of FeNi. Coupling constants corresponding to Fe-Fe, Fe-Ni, and Ni-Ni pairs are shown by black, blue, and green markers, respectively. (b) Ferromagnetic Curie temperature $T_{\text{c}}$, obtained from Monte Carlo simulations, as a function of the cutoff distance used for the Fe-Fe interactions (see text).}
   \label{img:Tc_vs_distance}
\end{figure}

From Fig.~\ref{img:J_vs_d_comparison_Alex}(a) one can see that there are noticeable Fe-Fe couplings even for rather large atomic distances. To further analyze this distance dependence, Fig.~\ref{img:Tc_vs_distance}(a) shows the same coupling constants as in Fig.~\ref{img:J_vs_d_comparison_Alex}, but multiplied with the cube of the corresponding inter-atomic distances, i.e. $J_{ij} \cdot d_{ij}^3$. For the Fe-Ni and Ni-Ni couplings, the corresponding data-points still converge quickly towards zero for large distances, which means that these couplings decay with distance faster than $d_{ij}^{-3}$. On the other hand, for the Fe-Fe pairs, one can see that the data-points oscillate and do not seem to decay even for very long atomic distances.
As pointed out in previous works~\cite{Alex, Belozerov_2020}, this indicates an approximate $d_{ij}^{-3}$ dependence of the Fe-Fe interaction which, together with the oscillatory behavior, is typical for metals with RKKY-like exchange interactions.

Fig.~\ref{img:Tc_vs_distance}(b) shows the ferromagnetic Curie temperature, $T_{\text{c}}$, obtained from Monte Carlo simulations of the Heisenberg model, where the Fe-Fe interactions are considered only up to a certain maximum inter-atomic distance. For the Fe-Ni and Ni-Ni pairs, all calculated coupling constants have been included, i.e., up to a very large distance of $10 a_0$. Note that due to the fast decay of these couplings, the following results should be unaffected by the specific cutoff-distance used for the Fe-Ni and Ni-Ni interactions.
It can be seen that, due to the long range of the Fe-Fe interaction, the calculated $T_c$ exhibits strong variations as function of the Fe-Fe cutoff distance, but seems to converge to a value around 700\,K once all Fe-Fe interactions up to a distance of around 8\,\AA{} are taken into account. This shows that, in order to obtain a reliable estimate of $T_{\text{c}}$ in this system, it is essential to include Fe-Fe interactions up to rather large distances.   

Based on these test calculations, we now obtain an accurate estimate for $T_{\text{c}}$ from Monte Carlo simulations of the Heisenberg model including all calculated coupling constants up to a maximum distance of $10 a_0$ and then perform a Binder cumulant analysis, as described in Sec.~\ref{sec:MFT}. We obtain a value of $T_c = 736$\,K. 
Note that for the test calculations shown in Fig.~\ref{img:Tc_vs_distance}(b), $T_{\text{c}}$ is obtained simply from the peak position of the calculated temperature dependence of the specific heat, and thus differs somewhat from the more accurate value obtained via the Binder cumulants.
Our calculated $T_{\text{c}}$ agrees well with the value of about 780\,K obtained by Tian \textit{et al.}~\cite{Tian2020}, whereas the $T_{\text{c}}$
of 916\,K obtained by Edstr\"om \textit{et al.}~\cite{Alex} is noticeably higher.
This is due to the stronger second nearest neighbor Fe-Fe and first nearest neighbor Fe-Ni coupling constants obtained in Ref.~\onlinecite{Alex} (see Fig.~\ref{img:J_vs_d_comparison_Alex}).
Note that both our and the value of Tian~\textit{et al.} are lower than the experimental $T_c$ of the disordered system and thus seem to underestimate the ``true'' Curie temperature of the ordered state.

\subsection{\label{subsec:couplings_disordered}Magnetic interactions for (partially) disordered FeNi}

\begin{figure*}
    \includegraphics[width=0.9\textwidth]{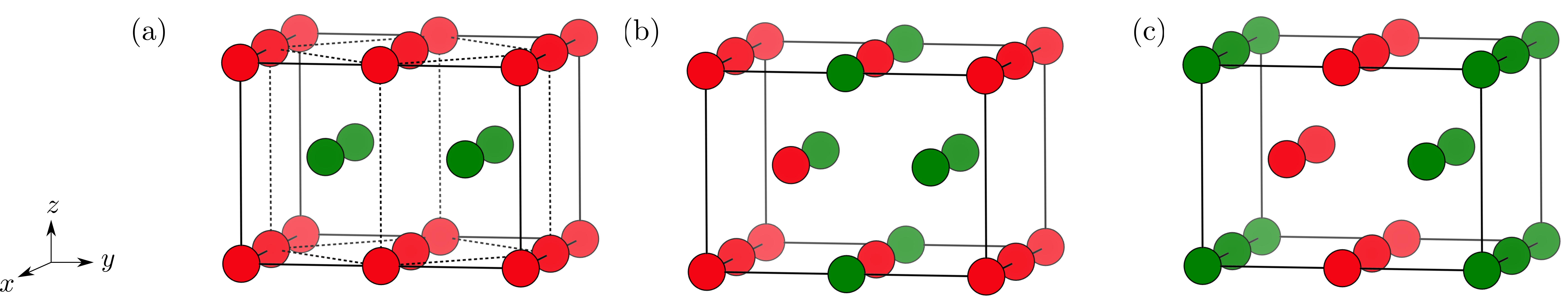}
    \caption{Different ordered and (partially) disordered configurations considered in this work, depicted in the $2 \times 2 \times 1$ supercell relative to the primitive tetragonal cell of the L$1_0$ structure: (a) ordered, (b) 1-pair-exchanged, and (c) 2-pairs-exchanged. Fe and Ni atoms are represented by red and green spheres, respectively. Dotted lines in (a) indicate the underlying conventional cubic cell of the fcc lattice.}
    \label{img:disordered_supercells}
\end{figure*}

We now investigate the effect of chemical disorder on the magnetic exchange couplings by starting from the fully ordered case, and then successively exchanging the positions of one or two pairs of Fe and Ni atoms within an 8-atom supercell, resulting in the two configurations shown in Fig.~\ref{img:disordered_supercells}(b) and (c).
Note that all other configurations that can be created by exchanging one Fe-Ni pair in this 8-atom supercell are equivalent to the one shown in Fig.~\ref{img:disordered_supercells}(b), whereas several distinct configurations can be created by exchanging two Fe-Ni pairs. For simplicity we limit our study to the configuration depicted in Fig.~\ref{img:disordered_supercells}(c).
In the following, we refer to these two configurations as ``1-pair-exchanged'' and ``2-pairs-exchanged'', respectively. 

\begin{figure*}
    \includegraphics[width=1.0\textwidth]{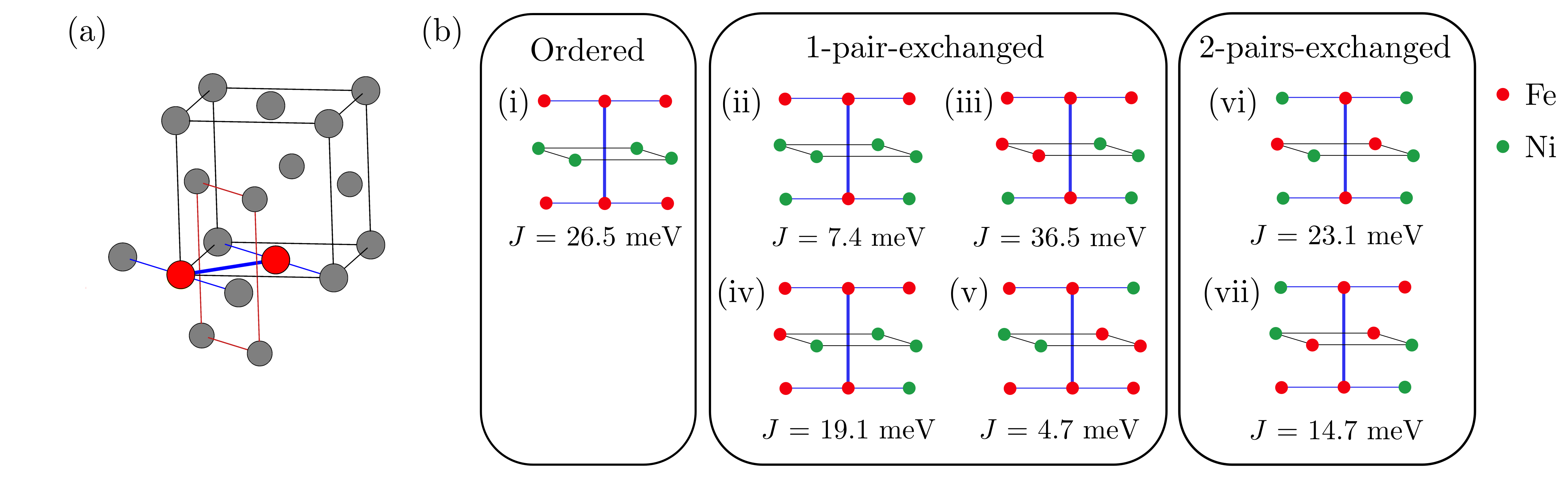}
    \caption{(a) Depiction of the shared first nearest and first/second nearest neighbor environment for a pair of nearest neighbor sites (indicated as red colored spheres) in an fcc lattice. The conventional cubic cell is also shown for clarity. (b) Schematic showing the occupation of the shared first and first/second nearest neighbour environments for all nearest neighbor Fe-Fe pairs that are present in the ordered [(i)], 1-pair-exchanged [(ii)-(v)], and 2-pairs-exchanged configurations [(vi)-(vii)] along with the corresponding values of the magnetic coupling constant $J$.}
    \label{img:J_env_corr}
\end{figure*}

In Fig.~\ref{img:J_env_corr}(b), we list the magnetic exchange couplings obtained for all inequivalent first nearest neighbor Fe-Fe pairs in the three different configurations. The corresponding local atomic environments are also indicated and will be discussed further below.
One can see that the calculated values vary drastically, from 4.7\,meV to 36.5\,meV, while the corresponding value in the fully ordered structure is 26.5\,meV.
We note that, if the nearest neighbor Fe-Fe coupling would be completely configuration-independent, then all values listed in Fig.~\ref{img:J_env_corr}(b) would be identical. 
The large variation of the Fe-Fe nearest neighbor coupling in the different cases thus shows that the local chemical environment has a significant influence on the magnitude of the magnetic exchange interactions in FeNi.
As already discussed in Sec.~\ref{sec:Intro}, such variations are not captured by effective medium methods such as the CPA, which are often used to model chemical disorder in alloys.

\begin{table}
\caption{\label{tab:table2}%
First nearest-neighbour Fe-Fe magnetic exchange interaction (in meV) for relaxed and unrelaxed 1-pair-exchanged and 2-pairs-exchanged configurations.}
\begin{ruledtabular}
\renewcommand{\arraystretch}{1.2}
\begin{tabular}{@{}lccc@{}}
% Supercell & $J$ &\\
&  Unrelaxed & Relaxed\\
\hline
1-pair-exchanged 
 & 7.4 & 10.1 \\
 & 36.5 & 38.8\\
 & 19.1 & 19.9 \\
 & 4.7 & 6.8 \\
\hline
2-pairs-exchanged & 23.1 & 21.9\\
 & 14.7 & 15.5\\
\end{tabular}
\end{ruledtabular}
\end{table}

The magnetic exchange couplings shown in Fig.~\ref{img:J_env_corr}(b) are obtained by decorating the sites within a perfect fcc lattice in different ways with Fe and Ni atoms, without allowing the atomic positions to relax within the resulting lower symmetry. In order to assess the effect of such relaxations, we now recalculate the magnetic coupling constants for the 1-pair-exchanged and 2-pairs-exchanged configurations after allowing all atomic positions to relax, while still keeping the lattice vectors of the supercell fixed.
The results for the first nearest neighbor Fe-Fe couplings are shown in Table~\ref{tab:table2} and are compared to the corresponding values for the unrelaxed case. It can be seen that the relaxation leads to changes in the magnetic coupling constants of up to about 2\,meV, but the effect is clearly significantly weaker than the effect due to the different chemical environments.

The strong configuration dependence of the first nearest Fe-Fe coupling raises the question of whether it is possible to identify simple rules on how the strength of this coupling depends on the local atomic environment. For this purpose, we analyze the distribution of atoms on the sites that are closest neighbors to \emph{both} Fe atoms forming the pair under consideration. As shown in Fig.~\ref{img:J_env_corr}(a) there are four such sites, which form a square in the mid-plane perpendicular to the line connecting the two coupled sites. These four sites represent the minimal environment to be considered in any model describing the configuration dependence of the first nearest neighbor coupling. 
The next ``shell'' around the coupled Fe-Fe pair is formed by those sites that are first nearest neighbors to one of the coupled sites and second nearest neighbors to the other site. There are again four such sites, which form a rectangle in the plane parallel to the Fe-Fe distance vector and perpendicular to the plane formed by the common first nearest neighbors (see Fig.~\ref{img:J_env_corr}(a)).

In Fig.~\ref{img:J_env_corr}, we schematically depict the occupation of both the shared first nearest and the shared first/second nearest neighbor sites for all the inequivalent nearest neighbor Fe-Fe pairs included in the ordered, 1-pair-exchanged, and 2-pairs-exchanged configurations, along with the value of the corresponding magnetic exchange couplings. 
It is obvious that the shared first nearest neighbor environment is not sufficient to classify the different coupling constants, since, e.g., cases (iii), (v), and (vi) all have an equivalent shared first nearest neighbor environment but exhibit vastly different magnetic coupling constants (including both the highest and lowest calculated values of 36.5\,meV and 4.7\,meV). The same holds for cases (i) and (ii).

Considering both the shared first and first/second nearest neighbor environment, all inequivalent Fe-Fe pairs contained in our three configurations exhibit different local environments, which is in principle compatible with a local model for the exchange coupling based on this environment. However, to really establish or disprove such a model, one has to consider much larger supercells, that allow to sample more configurations, and also include cases with identical first/second nearest neighbor environment but different further neighbor environment. This would require an excessive computational effort. Considering that, in general, the applicability of a short-range local model for an itinerant magnetic system such as FeNi is rather questionable, we therefore refrain from sampling further couplings using larger and larger supercells.
Instead, we try to estimate the effect of a strong configuration dependence of the magnetic coupling constants on the Curie temperature of a disordered magnetic system using a simple Heisenberg model with random couplings.

\subsection{\label{subsec:model_study}Model study with random couplings}

The very high sensitivity of the  magnetic coupling constants on the specific chemical environment, and the high computational effort to fully resolve this configuration dependence (if at all possible), represents a big obstacle for the reliable estimation of magnetic ordering temperatures for disordered itinerant magnets such as FeNi from first principles calculations.
In the following, we therefore employ a strongly simplified model to obtain a rough estimate of how the strong configuration-dependent variations of the exchange couplings can affect the Curie temperature of a disordered magnetic system, in comparison to the Curie temperature obtained using only configuration-independent ``average'' magnetic coupling constants. 

Specifically, we consider a Heisenberg model with only nearest neighbor interactions on an fcc lattice, and we approximate the configuration-dependent variations of the magnetic coupling constants by a Gaussian-distributed random variable, where the mean value of the Gaussian distribution represents the average coupling constant, and its standard deviation, $\sigma$, quantifies the configuration-dependent variations. We then perform temperature-dependent Monte Carlo simulations as outlined in Sec.~\ref{sec:MFT}, and analyze how the obtained Curie temperature depends on $\sigma$, i.e., on the strength of the variation in the magnetic coupling constants.

\begin{figure}
   \centering
   \includegraphics[width=0.5\textwidth]{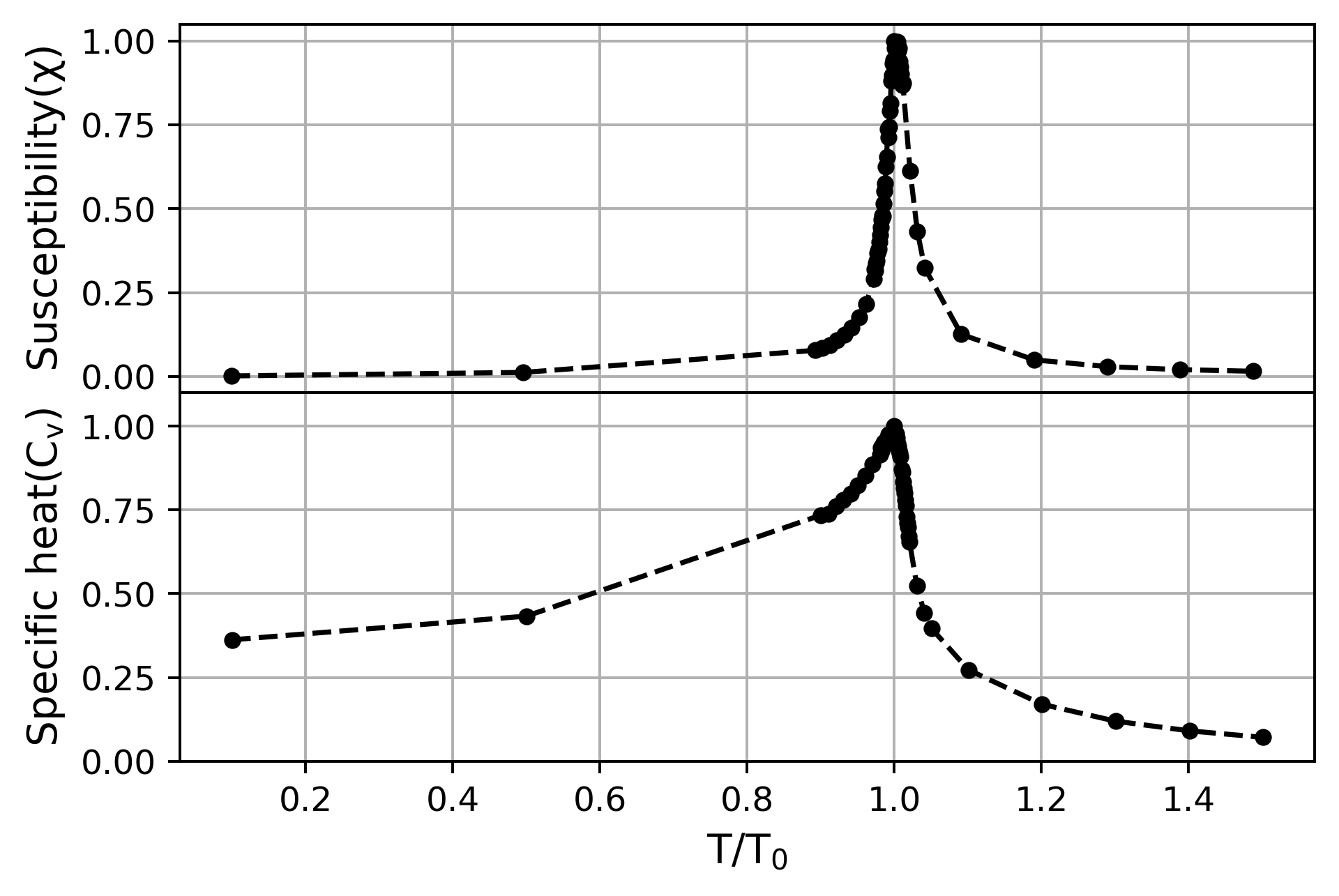}
   \caption{(a) Susceptibility, and (b) specific heat as a function of temperature (in relative units) obtained from Monte Carlo simulations for a simple Heisenberg model with only nearest neighbor couplings on an fcc lattice for the case when all coupling constants are identical ($\sigma=0$). $T_0$ is the critical temperature for $\sigma=0$.}
   \label{img:Cv_X_comparison}
\end{figure}

Fig.~\ref{img:Cv_X_comparison} shows the calculated susceptibility ($\chi$) and specific heat ($C_v$) as a function of temperature for this model for the case when all coupling constants are identical to the average one ($\sigma = 0$). One can see that both $\chi$ and $C_v$ exhibit clear peaks at the same critical temperature $T_0$. However, since the peak in the susceptibility appears much sharper than the one in the specific heat, in the following we use the peak value of $\chi$ to accurately determine the Curie temperature as function of $\sigma$.

\begin{figure}
   \centering
   \includegraphics[width=0.5\textwidth]{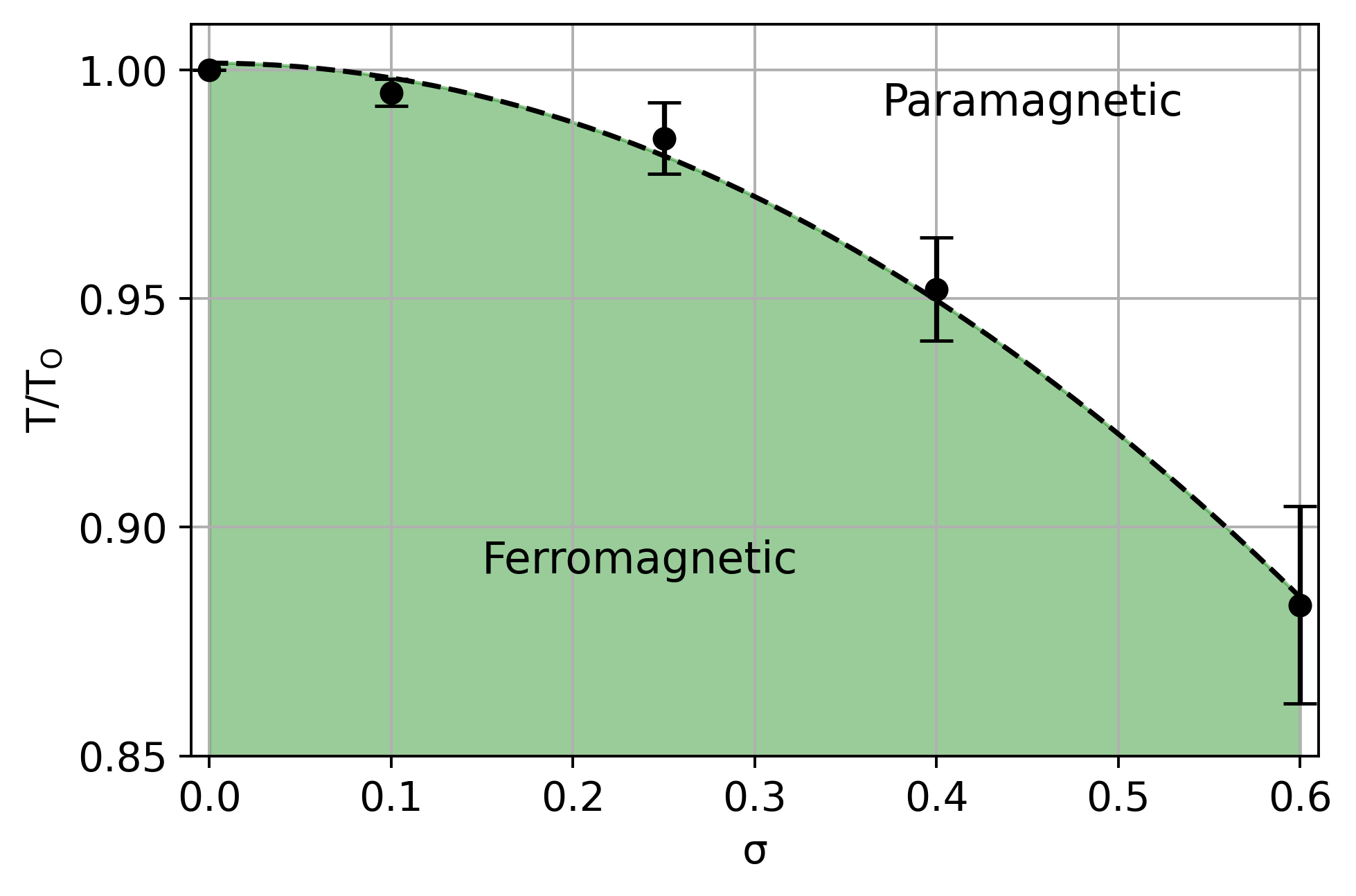}
   \caption{Curie temperature $T_c$ of the Heisenberg model with Gaussian-distributed coupling constants as a function of the standard deviation, $\sigma$ (defined relative to the mean value of the Gaussian distribution). $T_0 = T_c(\sigma = 0)$ is the Curie temperature obtained for the average coupling, and the error bars indicate the standard deviation of the mean obtained by averaging over 100 instances with different random seeds for each $\sigma$ (see text). The black dashed line is a quadratic fit to the data.
   }
   \label{fig:Tc_vs_sigma}
\end{figure}

Fig.~\ref{fig:Tc_vs_sigma} shows the variation of the average Curie temperature, $T_{\text{c}}$, relative to $T_0$, as a function of the standard deviation $\sigma$ of the Gaussian distributed random magnetic coupling constants. Note that $\sigma$ corresponds to a Gaussian distribution with a mean equal to the average magnetic coupling, and is therefore defined relative to this average coupling.
One can see that $T_{\text{c}}$ decreases with increasing standard deviation $\sigma$, i.e. with increasing ``randomness'' of the magnetic exchange couplings, and that this decrease can be fitted well with a quadratic dependence.
This indicates that using configuration-independent ``average'' magnetic exchange couplings obtained by effective medium approaches such as CPA are expected to overestimate the critical temperature of a disordered system.

If we take the seven different values for the configuration-dependent nearest neighbor interaction obtained from the ordered, the 1-pair-exchanged, and the 2-pairs exchanged configuration and evaluate the empirical standard deviation, we obtain $\sigma \approx 0.6$ (relative to an average value of 18.9\,meV). Comparing this with the data shown in Fig.~\ref{fig:Tc_vs_sigma}, this would correspond to a reduction of $T_c$ of about 10\,\% compared to the value obtained using only an average coupling strength.
We note that this is clearly a very naive estimation based on a very small number of samples, but it shows that a strong configuration dependence of the magnetic coupling, as observed for FeNi in Sec.~\ref{subsec:couplings_disordered}, can indeed lead to a noticeable reduction of $T_c$ compared to that obtained from average effective medium couplings. 

Note that the effect of random magnetic exchange couplings on the magnetic transition temperature of a Heisenberg model with nearest and next-nearest antiferromagnetic coupling on a two-dimensional square lattice was studied previously by Li \textit{et al.}~\cite{Li_2015}. Thereby, the nearest neighbor coupling was obtained from a homogeneous random distribution within a given range, and it was also found that the transition temperature decreases with increasing variation of the random magnetic exchange couplings.

\section{\label{sec:Summary}Summary and Conclusions}

In summary, we have investigated the effect of chemical disorder on the magnetic exchange couplings in L1$_0$-FeNi using first principles DFT calculations.
Thereby, we have used supercells with different atomic distributions, to specifically include effects due to the symmetry-broken local environments around the individual atoms that are not included in effective medium approaches such as CPA. 
We find that such effects can lead to rather large variations of the magnetic exchange couplings, exemplified by our analysis of the nearest neighbor Fe-Fe coupling, which exhibits values ranging from 4.7\,meV up to 36.5\,meV, with a value of 26.5\,meV obtained for the fully ordered structure. 

Our analysis of the shared first and shared first/second nearest neighbor chemical environments of the different inequivalent Fe-Fe nearest neighbor pairs included in our supercells indicates that the strength of the couplings is affected by long range effects that go beyond the closest neighbor shell, as can be expected for an itinerant metallic material such as FeNi.
The lack of a simple relation that relates the variation of an individual $J_{ij}$ (relative to the corresponding average value) to its local chemical environment makes it very challenging to incorporate effects beyond CPA in the calculation of $T_{\text{c}}$. 
However, our simple model study using random Gaussian-distributed nearest neighbor couplings on an fcc lattice suggests that such local variations of the coupling constants can lead to a reduction of $T_{\text{c}}$ of up to 10\,\%, compared to that obtained using only an average coupling.

Thus, there is a hierarchy of effects that in general tend to reduce the magnetic ordering temperature in random alloys such as partially ordered FeNi. First, the ``average'' coupling strength is affected by the chemical disorder. For example, Tian \textit{et al.} obtain a reduction of around 30-35\,\% of the nearest neighbor Fe-Fe coupling obtained within CPA for the disordered system compared to the fully ordered case~\cite{Tian2020}. This is in principle consistent with our supercell calculations, where the simple average of this coupling over all inequivalent Fe-Fe pairs in our two disordered configurations gives about 17.6 meV, i.e., a reduction by about 35\,\% compared to the ordered case. However, one should note that this average is based on only very few samples.
Second, the random connectivity between different magnetic atoms (Fe and Ni in our case) will also affect the Curie temperature relative to the ordered case, which corresponds to a very regular network of Fe-Fe, Fe-Ni and Ni-Ni bonds.
Finally, the variation of the coupling strength according to the specific chemical environment around the atoms in the random alloy can lead to a further decrease of $T_{\text{c}}$, as indicated by our simple model. It is mainly this last effect that we have quantified within this work.

Furthermore, it appears that our first-principles-based results are underestimating the (hypothetical) Curie temperature of the fully ordered system. It is unclear whether this underestimation is related to the general applicability of the (classical) Heisenberg model to FeNi or whether it is caused by deficiencies of the generalized gradient approximation in the underlying DFT calculations (or other approximations in the method). However, we note that the comparison with the results obtained by Edstr\"om \textit{et al.}~\cite{Alex} also demonstrates that moderate changes in specific calculated coupling constants can lead to rather strong differences in the predicted Curie temperatures.  

\begin{acknowledgments}
We are grateful to Alexander Edstr\"om for many helpful discussions and for providing the data from Ref.~\onlinecite{Alex} included in Fig.~\ref{img:J_vs_d_comparison_Alex}.
We also thank Maximilian Merkel and Alberto Carta for help with using the {\tt wannier90} code. This work was supported by ETH Z\"urich. Calculations were performed on the cluster \enquote{Piz Daint}, hosted by the Swiss National Supercomputing Centre, and the \enquote{Euler} cluster of ETH Z\"urich.
\end{acknowledgments}

\appendix

\section{Orbital decomposition of nearest neighbor Fe-Fe coupling}

Eq.~\eqref{eq:Jij_MFT} in principle allows for a decomposition of the coupling constants $J_{ij}$ into different orbital contributions. However, for the case of L1$_0$-FeNi, the quantities appearing in Eq.~\eqref{eq:Jij_MFT}, in particular the exchange splitting $\Delta_i^{mm'}$, contain off-diagonal elements mixing the $d$ and $s$ type Wannier orbitals. Nevertheless, by restricting the summation in Eq.~\eqref{eq:Jij_MFT} to only the diagonal elements $\Delta_i^{mm}$ and considering only contributions from the $d$-type orbitals, one can obtain a ``$d$-only'' contribution to the magnetic coupling. 

In this way, we obtain a $d$-only contribution to the Fe-Fe coupling of 29.6\,meV (second row in Table~\ref{tab:Jij_2atomcell}). This value is larger than the full value of 27.2\,meV (first row in Table~\ref{tab:Jij_2atomcell}), obtained by considering all contributions in Eq.~\eqref{eq:Jij_MFT}, showing that here the combination of $sp$ and orbitally mixed terms leads to a small negative contribution to $J_{ij}$~\footnote{The small difference to the value of 26.5\,meV listed in Fig.~\ref{img:J_env_corr} is due to the the different unit cells (8-atom cell versus 2-atom cell) used in the calculations.}. Furthermore, it demonstrates that, as probably expected, the main contribution to the magnetic coupling stems from the $d$ orbitals.

\begin{table}[t]
\caption{\label{tab:Jij_2atomcell}%
Magnetic exchange couplings, $J_{ij}$, for the nearest-neighbor Fe-Fe interaction in L1$_0$-FeNi obtained using different sets and subsets of Wannier functions. The first column indicates the different sets of Wannier functions that have been constructed, while the second column indicates which terms are considered in Eq.~\ref{eq:Jij_MFT} when evaluating the corresponding $J_{ij}$. All calculations are performed for the primitive 2-atom unit cell.}
\begin{ruledtabular}
\renewcommand{\arraystretch}{1.2}
\begin{tabular}{@{}lcc@{}}
Wannier set & Terms in Eq.~\ref{eq:Jij_MFT} & $J_{ij}$ (meV) \\
\hline
$d+s+p$ (projections) & all  &27.2 \\
$d+s+p$ (projections) & $d$-only & 29.6 \\
$d$ (projections) & all & 35.7\\
$d$ (MLWFs) & all & 43.4 \\
\end{tabular}
\end{ruledtabular}
\end{table}

This raises the question of whether it would be sufficient to consider only the $d$-bands in the first place, i.e., construct a smaller set of Wannier functions describing only the $d$ bands, and still obtain a good estimate for $J_{ij}$. To test this hypothesis, we construct two additional sets of Wannier functions, where we include only five $d$-orbitals per atom. For the first set, we obtain the Wannier functions from orthonormalized atomic projections as before, while for the second set we perform a subsequent minimization of the quadratic spread functional to obtain MLWFs.
In both cases we use an (outer) energy window ranging from $-10$\,eV to about 5\,eV above the Fermi level and obtain a set of atom-centered $d$-like orbitals suitable to evaluate Eq.~\eqref{eq:Jij_MFT}. In both cases also the Wannier-interpolated bands resemble the DFT band-structure in the energy range of the $d$-bands, i.e., between approximately $-5$\,meV and the Fermi level for the majority spin channel. 

The magnetic coupling constants for the nearest-neighbour Fe-Fe coupling obtained from these two additional sets of Wannier functions are listed in the third and fourth row of Table.~\ref{tab:Jij_2atomcell}. 
It can be seen that the corresponding values (in particular for the set of MLWFs) are significantly larger than the $d$-only contribution obtained from the full description using also $s$ and $p$ bands.
This shows that, even though the $d$-orbitals make up the main contribution to the magnetic coupling constants, it is nevertheless important to include $s$ and $p$ states to accurately account for their effect on the $d$ band dispersion. We note that, due to the entanglement of $d$ and $sp$ contributions in the bandstructure of FeNi, the $d$ subset of the full $d+s+p$ Wannier basis and the two different $d$-only Wannier sets (projected and MLWFs) are all describing slightly different subspaces of the occupied Kohn-Sham states. 

Only the $d+s+p$ Wannier set results is a complete and accurate description of all occupied bands in FeNi, and therefore only the corresponding value of $J_{ij}$ should be considered as ``correct'' (or most accurate). Nevertheless, our analysis raises the question of a potential basis set dependence of the magnetic coupling constants, for example in cases where a complete description of all occupied bands can be achieved using different sets of Wannier functions, e.g., corresponding to different degrees of localization. In the present case, a more systematic analysis is hindered by the strong entanglement of bands and the fact that spread minimization on the full set of Wannier functions leads to orbitals that are not atom-centered. We therefore leave this question open for future research.

\bibliography{main}
\end{document}